# Competition between ferromagnetism and spin glass: the key for large magnetoresistance in oxygen deficient perovskites SrCo$_{1-x}$M$_x$O$_{3-\delta}$ (*M* = Nb, Ru)

T. Motohashi*, V. Caignaert, V. Pralong, M. Hervieu, A. Maignan, and B. Raveau

Laboratoire CRISMAT, UMR CNRS ENSICAEN 6508, 6 bd Maréchal Juin

14050 CAEN Cedex 4 France

(Dated: 28 January 2004)

The magnetic and magnetotransport properties of the oxygen deficient perovskites, SrCo$_{1-x}$M$_x$O$_{3-\delta}$ with *M* = Nb and Ru, were investigated. Both Nb- and Ru-substituted cobaltites are weak ferromagnets, with transition temperatures $T_m$ of 130-150 K and 130-180 K, respectively, and both exhibit a spin glass behavior at temperatures below $T_f$ = 80-90 K. It is demonstrated that there exists a strong competition between ferromagnetism and spin glass state, where Co$^{4+}$ induces ferromagnetism, whereas Nb or Ru substitution at the cobalt sites induces magnetic disorder, and this particular magnetic behavior is the origin of large negative magnetoresistance of these oxides, reaching up to 30% at 5 K in 7 T. The differences between Nb- and Ru-substituted cobaltites are discussed on the basis of the different electronic configuration of niobium and ruthenium cations.

*Corresponding author
Permanent address: Materials and Structures Laboratory, Tokyo Institute of Technology, 4259 Nagatsuta, Midori-ku, Yokohama 226-8503, Japan
E-mail: t-mot@msl.titech.ac.jp
Phone: +81-45-924-5318
Fax: +81-45-924-5339

# I. INTRODUCTION

Since the discovery of "colossal" magnetoresistance (MR) effect in perovskite manganites [1], search for next MR materials has been widely conducted in various transition-metal oxides. Among them, cobaltites with the perovskite structure are of potential interest for the discovery of magnetoresistive properties. In these oxides, cobalt is susceptible to accommodate various oxidation states ($Co^{2+}$, $Co^{3+}$, and $Co^{4+}$) and various kinds of magnetic states are competing with a subtle balance due to the variety of spin states and spin interactions of cobalt ions. Therefore, a dramatic modification in the magnetic state and thereby the MR property can be expected by tuning the chemical composition. Nevertheless, very few cobalt perovskites were found to exhibit a large MR effect. The oxygen deficient perovskite $SrCoO_{2.75}$ [2-4], prepared by a two step method, was found to show a negligible MR value of 0.5% in 7 T [4], whereas the MR properties were reported for neither Sr-rich $La_{1-x}Sr_xCoO_{3-\delta}$ ($0.5 \leq x \leq 0.9$) [5] nor the stoichiometric perovskite $SrCoO_3$ prepared by electrochemical oxidation [6] or under high pressure [7]. In fact, a relatively large MR effect was observed in two series of perovskites: La-rich $La_{1-x}Sr_xCoO_3$ ($x < 0.5$) [8-10] and the ordered oxygen-deficient perovskite $LnBaCo_2O_{5.4}$ ($Ln$: rare-earth elements) [11] whose MR value, defined as $-(\rho_H - \rho_0) / \rho_0$, ranges from 12.5% to 50% in 7 T at low temperatures.

Recently, a MR value of 12% in 7 T at 5 K was reported for the Sc-substituted oxygen-deficient perovskite $SrCo_{1-x}Sc_xO_{3-\delta}$ [12]. After this report, we successfully stabilized the perovskite cobaltite by introduction of niobium on the Co sites and found that the Nb-substituted cobaltite, $SrCo_{1-x}Nb_xO_{3-\delta}$, exhibits a large MR value of 30% at low temperatures [13]. This suggests that the appearance of large MR is the consequence of the modification of magnetism through Nb- or Sc-for-Co substitution. Thus, a detailed investigation on magnetism of such substituted perovskites is absolutely necessary to elucidate the origin of the large MR effect.

In the present paper, we report on the magnetic and magnetotransport properties of two series of oxygen deficient perovskites, $SrCo_{1-x}M_xO_{3-\delta}$ with $M$ = Nb and Ru. We show that these cobaltites exhibit a quite unusual magnetic behavior, i.e. weak ferromagnetism with transition temperatures $T_m$ of 130-150 K and 130-180 K for the Nb- and Ru-substituted phases, respectively, and a spin-glass like behavior at lower temperatures below 80-90 K. We explain the large MR effect of these phases on the basis of the competition between ferromagnetism and spin glass state, where $Co^{4+}$ induces ferromagnetism, whereas Nb or Ru substitution at the cobalt sites induces magnetic disorder. The differences between Nb and Ru are also discussed taking into account the difference in the electronic configuration of non-magnetic Nb ($Nb^{5+}$: $d^0$) and magnetic Ru ($Ru^{4+}$: $d^4$ or $Ru^{5+}$: $d^3$) cations.

## II. EXPERIMENTAL

Samples of $SrCo_{1-x}M_xO_{3-\delta}$ ($M$ = Nb and Ru) with $x$ = 0.05, 0.10, 0.15, and 0.20 were prepared by a conventional solid-state reaction. Powder mixtures of $SrCO_3$, $Co_3O_4$, $Nb_2O_5$, and $RuO_2$ with appropriate ratios were calcined in flowing $O_2$ at 800°C for 12 h. This calcined powder was ground, pelletized, and fired in flowing $O_2$ at 1200°C for 12 h, followed by slow cooling down to room temperature. A part of the samples was subsequently post-annealed at 500°C for 24 h in high oxygen pressure (10 MPa).

X-ray powder diffraction (XRPD) analysis was performed using a Philips diffractometer with Cu$K_\alpha$ radiation. The data were collected in an angular range of 20° ≤ 2θ ≤ 130° and analyzed with a Rietveld program, FULLPROF [14]. The electron diffraction (ED) and the energy dispersive spectroscopy (EDS) analyses were performed using a TEM, JEOL 200 CX equipped with a Kevex analyzer. The oxygen content, 3-δ, was determined by the iodometric titration. Details in the titration technique are given elsewhere [15].

Magnetic properties were investigated by means of a SQUID magnetometer (Quantum Design: MPMS) equipped with an ac-susceptibility option. In dc-magnetization (*M*) measurements, the samples were first zero-field cooled (ZFC) or field cooled (FC), and then *M*(*T*) was measured at increasing temperature. The *M*(*H*) measurements were also performed in a field up to 5 T. The ac-susceptibility (χ) was measured in an ac-field of 3 Oe with frequencies of 1, 10, and 100 Hz. Magnetotransport properties were studied with a four-points method using a PPMS facility (Quantum Design) in a temperature range of 5-300 K and in a magnetic field of 0-7 T. The MR vs. *H* measurements were performed at several temperatures. In each measurement, the samples were first zero-field cooled and then the applied field was increased from 0 to 7 T and decreased back to 0 T, and further to -7 T.

## III. RESULTS

### A. Chemical composition and lattice parameters

Single phased samples with the perovskite structure have been synthesized for $0.05 \leq x \leq 0.20$ for both cobaltites $SrCo_{1-x}Nb_xO_{3-\delta}$ and $SrCo_{1-x}Ru_xO_{3-\delta}$. Nevertheless the $x = 0.05$ samples are not stable and decompose after annealing in an oxygen pressure. The X-ray diffraction patterns of samples with high Nb and Ru contents ($x \geq 0.15$) are characteristic of a cubic-like perovskite unit cell. However, the electron diffraction analysis evidences that all the title compounds exhibit a tetragonal cell with $a \approx a_p$ and $c \approx 2a_p$ (the subscript "p" denotes the perovskite subcell). This observation is in agreement with the XRPD patterns of the Ru-substituted $x = 0.05$ and 0.10 samples, in which weak extra peaks are seen due to a tetragonal distortion (Fig. 1). There is no condition limiting of the reflection, involving $P4$, $P\bar{4}$, $P4/m$, $P422$, and $P4/mmm$ as possible space groups. A typical [010] electron diffraction pattern is shown in the inset of Fig. 1. Such a pattern evidences the formation of twinning domains, which originate from the superstructure ($c \approx 2a_p$) and the parameter relationship, i.e. $c \approx 2a$. The weak reflections of the 90°-oriented domain are indicated by white arrows in the figure. Moreover, it is important to note that, in particular for lower substituted samples ($x \leq 0.10$), there exist weaker extra spots suggesting additional local ordering phenomena. A detailed structural study of these oxygen deficient perovskites will be published elsewhere. Nevertheless, it must be emphasized that the lattice parameters and the volume of the perovskite subcell (Table I) increase linearly with the Nb or Ru content $x$.

The EDS analysis showed that cations are distributed homogeneously and that the actual cationic composition is close to the nominal one. The chemical analysis, using the redox titration method, showed a significant deviation from the "$O_3$" stoichiometry (Table I), but the $\delta$ value does not vary dramatically with $x$, ranging from 0.17 to 0.26 for $M$ = Nb, and from 0.13 to 0.26 for $M$ = Ru. Note that the oxygen content of the Ru-substituted samples is much increased by $O_2$ annealing as compared to the Nb-substituted samples.

### B. Magnetic properties

The temperature dependence of magnetization ($M$) collected in a magnetic field of 0.3 T for both sample series, $SrCo_{1-x}Nb_xO_{3-\delta}$ and $SrCo_{1-x}Ru_xO_{3-\delta}$, shows a significant increase in $M$ as $T$ decreases (Fig. 2), indicative of the presence of ferromagnetic interactions. The $M(T)$ curves of both series are rather smooth so that the transition temperature $T_m$ cannot be determined accurately, $T_m$ ranging from 130 K to 180 K. In both series magnetization is significantly increased by $O_2$ annealing (in particular for the Ru-substituted samples), due to the increase in $Co^{4+}$ concentration. Such a large impact of $O_2$ annealing upon the enhancement in ferromagnetism has been previously reported for the non-substituted $SrCoO_{3-\delta}$ cobaltite [2]. The influence of the substitution level upon the magnetic properties is somewhat different between Nb and Ru. In the $SrCo_{1-x}Nb_xO_{3-\delta}$ series (Figs. 2a and 2b), the absolute value of $M$ rapidly decreases with increasing the Nb content $x$ (Table II), being consistent with the fact that the non-magnetic $Nb^{5+}$ cations dilute the ferromagnetic interactions. On the other hand, for the $SrCo_{1-x}Ru_xO_{3-\delta}$ series the $M$ value of the as-synthesized samples systematically decreases with increasing $x$ up to 0.15 and then increases again for $x = 0.20$ (Table II). In addition, for the $O_2$-annealed Ru-samples the shape of the $M(T)$ curve is different from the Nb-samples. This different behavior of Ru with respect to Nb, will be explained further by the magnetic nature of the ruthenium cations.

These first results clearly show that the ferromagnetism in these two series of substituted cobaltites is much weaker than that reported for the stoichiometric $SrCoO_3$ [6,7]. The $T_m$ value is indeed much lower than the Curie temperature, $T_C = 280$ K, observed for $SrCoO_3$ and the maximum $M$ value, reached for the $O_2$-annealed $x = 0.10$ Ru-sample (Table II) is only of 1.18 $\mu_B$/f.u. instead of 1.8-2.1 $\mu_B$/Co for $SrCoO_3$ [6,7]. Nevertheless, it is worth pointing out that this value is four times higher than for the Sc-substituted cobaltite (0.3 $\mu_B$ /f.u.) [12]. The particular nature of ferromagnetism in these cobaltites is confirmed in the $M(H)$ curves registered at 5 K as exemplified by the $SrCo_{1-x}Nb_xO_{3-\delta}$ series (Fig. 3). One indeed observes that, though the $M(H)$ loops are characteristic of a ferromagnet, the $M$ value at 5 K in 5 T is so far not saturated decreasing rapidly from 0.73 $\mu_B$/f.u. for $x = 0.05$ to 0.12 $\mu_B$/f.u for $x = 0.20$, due to the dilution effect induced by Nb-for-Co substitution.

The ac-susceptibility curves $\chi'(T)$ and $\chi''(T)$ of the two series (Figs. 4 and 5) shed light on their unusual magnetic behavior. As exemplified by the $SrCo_{1-x}Nb_xO_{3-\delta}$ series whose curves were recorded in an ac-field of 3 Oe with a frequency of 1 Hz (Fig. 4), a significant increase in $\chi''$ is observed, for instance around 150 K for $x = 0.05$, corresponding to the appearance of dissipation linked to ferromagnetism. Though it is difficult to accurately determine the magnetic transition temperature $T_m$ due to a gradual change in susceptibility with temperature, $T_m$ can be estimated as the temperature where the magnitude of $\chi''$ drops to 1% of the maximum value. The thus obtained values (Table II) confirm the results of the $M(T)$ measurements and show that $T_m$ does not depend strongly on the Nb or Ru content and it increases slightly by $O_2$ annealing. More importantly, one observes a bump at $\approx$ 80 K, as already observed for the Sc-substituted phase [12]. The ac-susceptibility curves of the $O_2$-annealed $SrCo_{0.9}M_{0.1}O_{3-\delta}$ samples with $M$ = Nb (Figs. 5a and 5b) and $M$ = Ru (Figs. 5c and 5d) clearly show that the peak that appears at 80 K and 90 K, respectively, is frequency-dependent and shifts toward higher temperature as frequency, $f$, increases from 1 to 100 Hz. Such a behavior has already been observed in numerous manganites such as $Pr_{0.5}Ca_{0.5}Mn_{1-x}M_xO_3$ doped at the Mn sites with various cations ($M$ = Al, Ti, In, Fe, Sn, and Ga) [16] and it corresponds to a disordered magnetic state characteristic of a spin glass or a cluster glass. The frequency shift, defined as $\Delta T_m / [T_m \Delta(\log f)]$ and determined at 0.02, being close to the value reported for typical spin glass materials [17], supports this statement.

In summary, the unusual magnetic behavior of the Nb- and Ru-substituted perovskite cobaltites can be described as the consequence of the competition between a ferromagnetic state which stems from the presence of tetravalent cobalt species and a spin-glass state induced by the presence of Nb or Ru impurities at the Co sites. It is this "mixed state" which may involve phase separation which is the origin of the appearance of large MR in those materials.

C. Magnetotransport properties

Bearing in mind our recent results about the large MR effect observed for the $SrCo_{1-x}Nb_xO_{3-\delta}$ phase [13], we have completed the latter study by the exploration of the magnetotransport properties of $SrCo_{1-x}Ru_xO_{3-\delta}$.

The temperature dependence of resistivity ($\rho$) for these two series of cobaltites (Fig. 6) shows a very similar behavior. In both series $\rho$ increases with the Nb or Ru content, due to the decrease in the number of conducting paths, interrupted by the Nb or Ru species. Nevertheless one can notice that the magnitude of $\rho$ does not increase so rapidly with $x$ in the Ru series (Figs. 7c and 7d) as compared to the Nb series (Figs. 7a and 7b). It results that for larger $x$ values ($x$ = 0.15-0.20) the Ru-substituted phase has much lower resistivity than the Nb-substituted one. It must be emphasized that for both series $\rho$ is decreased by $O_2$ annealing, in agreement with the increase in $Co^{4+}$ content. All the samples show a semiconductive upturn at low temperatures, which is enhanced as the Nb or Ru content increases. It is also remarkable that for small $x$ values ($x \leq 0.10$) the low-temperature resistivity of the Nb- and Ru-substituted phases is as low as that reported for the stoichiometric $SrCoO_3$ phase, $\rho = 1.1 \times 10^{-2}$ $\Omega$ cm at 5 K [6].

It can be seen that in both series the 7-T curve starts to deviate from the 0-T curve around 150 K (Fig. 6). Thus whatever the substituted element is, i.e. Nb or Ru, a large negative MR effect is observed, whose onset coincides with the magnetic transition temperature $T_m$. The MR($H$) measurements at several temperatures were already performed previously for $SrCo_{1-x}Nb_xO_{3-\delta}$ [13] and will not be detailed here. Note only that the MR value monotonically increases with decreasing temperature and that the highest MR value in 7 T reaches 30% at 5 K for the $O_2$-annealed $x$ = 0.15 and 0.20 samples, which is comparable to the value reported for the well known-MR material $Sr_2FeMoO_6$ [18]. The comparison of the MR($H$) curve of $SrCo_{0.85}Nb_{0.15}O_{3-\delta}$ with that of $SrCo_{0.85}Ru_{0.15}O_{3-\delta}$ (Fig. 7) shows a remarkable contrast: i.e. the $O_2$-annealed Ru-phase exhibits much smaller MR than the Nb-one, in spite of the same substitution level ($x$). The greater ability of niobium to induce a large MR effect, as compared to Ru, has been observed for all the other samples, as shown by the plot of the MR values (at 5 K in 7 T) against $x$ for the two series of $SrCo_{1-x}M_xO_{3-\delta}$ with $M$ = Nb and Ru (Fig. 8). One observes that the MR magnitude of the Nb-substituted samples increases with $x$ and remains at the highest values for high Nb contents, whereas for the Ru samples the MR magnitude is smaller and moreover tends to decrease when $x$ is larger than 0.10.

## IV. DISCUSSION

It was previously reported that the magnetic behavior of the perovskite cobaltite $SrCoO_{3-\delta}$ is very sensitive to oxygen nonstoichiometry or cobalt valency, $V_{Co}$ [2,5-7]. For instance, the ferromagnetic Curie temperature $T_C$ and saturation magnetization $M_s$ are systematically lowered with decreasing the oxygen content "$O_{3-\delta}$" (or $V_{Co}$) from 280 K and 2.1 $\mu_B$/Co for the stoichiometric $SrCoO_3$ ($V_{Co}$ = 4.0) [6] to 180 K and 0.67 $\mu_B$/Co for $SrCoO_{2.745}$ ($V_{Co}$ = 3.49) [2], respectively. In the present study, the $V_{Co}$ value for the $SrCo_{1-x}Nb_xO_{3-\delta}$ phase deduced from the chemical formula, supposing that niobium is pentavalent, ranges from 3.30 to 3.39 (Table I). For the $SrCo_{1-x}Ru_xO_{3-\delta}$ phase the determination of $V_{Co}$ from the chemical formula is not straightforward, since there are two possible valence states for ruthenium, i.e. $Ru^{4+}$ and $Ru^{5+}$. Nevertheless, from the chemical viewpoint in oxides (whatever they are), $Ru^{4+}$ is more easily oxidized into $Ru^{5+}$ for a very large molar ratio Sr (or Ba) / Ru (>> 1), especially when the oxide is prepared under oxygen. For this reason, we will admit [19] that ruthenium is pentavalent rather than tetravalent, leading to the $V_{Co}$ value ranging from 3.25 to 3.59 (Table I). In Fig. 9, we plot $T_C$ (or $T_m$) and $M$ at 5 K in 5 T for our $SrCo_{1-x}Nb_xO_{3-\delta}$ and $SrCo_{1-x}Ru_xO_{3-\delta}$ samples as a function of $V_{Co}$, together with data for the non-substituted $SrCoO_{3-\delta}$ phase in previous literatures. Although $T_C$ and $M$ increase with the $Co^{4+}$ content in these compounds, the evolution of the $M$-vs-$V_{Co}$ plot for the Nb- and Ru-substituted samples is clearly different from that observed for $SrCoO_{3-\delta}$. This indicates that the modification of the magnetic properties in the substituted samples cannot be explained only by the change in the cobalt valency.

Thus, it appears that the substituted Nb and Ru cations directly modify the nature of magnetic interactions. The Nb or Ru species do not modify significantly $T_m$, but suppress dramatically the $M$ value by a dilution effect, suggesting that the presence of Nb or Ru at the Co sites hinders the development of ferromagnetism and promotes the formation of ferromagnetic clusters below $T_m$. At lower temperatures below $T_f$ = 80-90 K, where the ac-susceptibility depends on frequency (see Fig. 5), a spin-glass or cluster-glass state appears due to disordering introduced by Nb or Ru cations and oxygen vacancies, while the ferromagnetic clusters still remain as demonstrated by the existence of residual magnetization in the $M(H)$ curves at 5 K. In other words, as already stated above, there exists a strong competition between the ferromagnetic state and the spin glass state in the whole temperature range, $T < T_m$. It is this strong competition which plays a crucial role in the appearance of the large MR effect. Such a behavior can be compared to the one observed for the $Pr_{1-x}Ca_xMnO_3$ manganites, in which the CMR effect takes place more easily in composition ranges where phase separation (PS) occurs due to a competition between two states [20,21], i.e. the vicinity of the boundary between ferromagnetic and orbital-charge ordered states ($x \approx 0.3$-$0.5$) [22,23] or between orbital-charge ordered state and cluster glass ($x \approx 0.8$-$0.9$) [23]. This similarity may imply the existence of the PS state also in the present substituted cobaltites. In fact, a neutron diffraction study on $La_{1-x}Ca_xCoO_3$ revealed [24] that a macroscopically inhomogeneous distribution of spin states, indicating the presence of the PS state, also exists in perovskite cobaltites.

Considering the above model, it is reasonable to admit that negative MR of the Nb- and Ru-substituted cobaltites stems from the reduction of spin-dependent scattering of carriers. For this reason the existence of ferromagnetism is an important factor for the appearance of the MR effect. The correlation between MR($H$) and $M$($H$) is consistent with this statement, and the enhancement of magnetization and accordingly of MR by $O_2$ annealing in the $x = 0.20$ Nb-sample [13] supports this viewpoint. No doubt that there is a threshold of the ferromagnetic fraction for the appearance of a large MR effect. However, a second factor should be also needed to induce large MR, i.e. a magnetic disorder due to the spin-glass state which increases the spin-dependent scattering of carriers and thus decreases electrical conductivity in the absence of magnetic field. At $T < T_f$, the degree of (ferromagnetic) spin alignment gets enhanced by applying magnetic fields and accordingly the spin-dependent scattering is significantly reduced, leading to a considerable negative MR effect. Note that this situation is different from that observed for $SrFe_{0.5}Co_{0.5}O_3$ [4] which exhibits a much smaller MR (~ 12%) and only a gradual increase in MR at lowering temperatures, in spite of a large magnetic moment ($M \approx 1.5$ $\mu_B$/f.u.). The absence of a spin glass state competing with the ferromagnetic state for the latter compound is in agreement with our model.

It is worthwhile to compare the magnetic and MR properties of this Sr-rich series $SrCo_{1-x}Nb_xO_{3-\delta}$ and La-rich series $La_{1-x}Sr_xCoO_3$ ($0 \leq x \leq 0.4$). Mahendiran and Raychaudhuri reported [10] that $La_{1-x}Sr_xCoO_3$ shows two distinct characteristic behaviors depending on the Sr content ($x$). (i) The compounds with $x \geq 0.2$ are ferromagnetic metals with only small MR values (4-8%) in the vicinity of $T_C$. (ii) The compounds with $x < 0.2$ are semiconductors which accompany a spin-glass like state and large MR at low temperatures. Interestingly, the $SrCo_{1-x}Nb_xO_{3-\delta}$ and $SrCo_{1-x}Ru_xO_3$ series exhibit great similarities with the latter compositions, i.e. $x < 0.20$, even though the cobalt valency is totally different. For instance, both $SrCo_{0.8}Nb_{0.2}O_{2.83}$ ($V_{Co} = +3.30$) and $La_{0.93}Sr_{0.07}CoO_3$ ($V_{Co} = +3.07$) show large MR in the whole temperature range below $T_m$, and a hysteresis or "memory" effect of MR and in both cases a disordered magnetic state seems to play a key role in the appearance of large MR. Clearly, both Nb- or Ru-for-Co substitution and oxygen vacancies highly promote a disordered magnetic nature, giving rise to excellent MR characteristics similar to Sr-poor compositions of $La_{1-x}Sr_xCoO_3$ ($x < 0.2$).

Let us finally discuss the different effects of Nb- and Ru-substitutions on the physical properties. As the number of substituted Ru atoms increases, the substituted Ru would have large effects on the spin alignment of the matrix. As suggested in Refs. 25 and 26, the interaction between Ru and Co spins is antiferromagnetic in the perovskite structure. Accordingly, instead of having a diluted disordered magnetic structure for the Nb-substituted phase, a net ferromagnetic state tends to be set by coupling antiferromagnetically the non-compensated $Co^{3+}$ (IS, $S = 1$ or HS, $S = 2$) and $Ru^{5+}$ ($S = 1/2$) magnetic moments. Furthermore, the relatively higher $T_m$ value of the $x = 0.15$ and 0.20 Ru-substituted samples may reflect a strong coupling of $Co^{3+}$–O–$Ru^{5+}$ bonds or may be due to the appearance of $Ru^{4+}$ species as $x$ increases. For the latter, a strong ferromagnetic interaction between $Ru^{4+}$ spins would occur, similarly to the $SrRuO_3$ ferromagnet [27,28].

Substituted Ru species are, unlike $Nb^{5+}$, able to participate in the carrier transport. We consider that through Ru substitution a Ru-derived "impurity band" is formed due to strong hybridization of the Ru $4d$ orbital. Such an impurity band gives an additional contribution to the electrical conduction, resulting in lower resistivity than for the Nb-substituted samples. The significant reduction of the MR effect can also be explained by considering the appearance of the Ru-derived band, since magnetotransport is very sensitive to the electronic structure near the Fermi level. We believe that the Ru-derived band deteriorates spin-dependent scattering of carriers due to its less correlated character, resulting in the suppressed MR effect in Ru-rich samples.

## V. CONCLUSIONS

The present study of the oxygen-deficient perovskite cobaltites $SrCo_{1-x}Nb_xO_{3-\delta}$ and $SrCo_{1-x}Ru_xO_3$ shows that these oxides are weak ferromagnets with transition temperatures $T_m$ of 130-150 K and 130-180 K, respectively, and exhibit a subsequent transition to a spin-glass state, at lower temperatures $T_f$ of 80-90 K. It demonstrates that a large magnetoresistance (MR) effect observed in these cobaltites, reaching up to 30% in 7 T at 5 K, is the result of the competition between the ferromagnetic state and the spin glass state, involving most likely phase separation. Thus, the spin glass state induced by Nb substitution hinders the development of ferromagnetism and lowers the electrical conductivity with respect to the ferromagnetic state and plays a capital role in the enhancement in MR.

The different effects of Ru and Nb substitutions can be explained by the fact that Ru spins directly interact with Co spins and modify the nature of magnetic interactions, whereas Nb cations solely dilute the ferromagnetic interactions. The suppression of the MR effect in Ru-rich samples can be explained by the appearance of a Ru-derived band which would deteriorate spin-dependent scattering of carriers.

We have seen that a disordered magnetic state induced by chemical substitutions is quite effective to enhance MR. This study also suggests that non-magnetic cations, like $Nb^{5+}$, are preferable for chemical substitutions at the magnetically active sites in order to introduce magnetic disorder, in contrast to magnetic cations which would modify the valence band structure and in some cases make a negative impact on the MR property, as demonstrated in the $SrCo_{1-x}Ru_xO_{3-\delta}$ phase. We believe that the present study shows a new strategy to search for MR materials in which magnetic dilution is an important ingredient to induce a large MR effect.

*Note added in proof.* After the submission of this manuscript, we have noticed that another study on perovskite cobaltites was published [29] in which the authors discussed an intergranular giant MR effect and a spontaneously phase separated state in $La_{1-x}Sr_xCoO_3$ with $x < 0.18$, based on small-angle neutron scattering data. Our results are very close and in perfect agreement with those reported in Ref. 29. Thus, the present study provides very important and supplementary information on the unique magnetic nature in perovskite cobaltites.

Table I:

The perovskite subcell parameter $a_p$, lattice parameters $a$ and $c$, volume of the perovskite subcell, oxygen content 3-$\delta$, and cobalt valency $V_{Co}$ of the $SrCo_{1-x}Nb_xO_{3-\delta}$ and $SrCo_{1-x}Ru_xO_{3-\delta}$ samples.

| Sample | | Subcell parameter $a_p$ (Å) | Lattice parameters | | Subcell volume (Å³) | Oxygen content 3-$\delta$ | Cobalt valency $V_{Co}$ |
| --- | --- | --- | --- | --- | --- | --- | --- |
| | | | $a$ (Å) | $c$ (Å) | | | |
| Nb-substituted ($x$) | 0.05 as-synthesized | 3.8560(1) | – | – | 57.466(2) | 2.74 | 3.39 |
| | 0.10 as-synthesized | 3.8666(1) | – | – | 57.806(2) | 2.74 | 3.32 |
| | 0.15 as-synthesized | 3.8739(1) | – | – | 58.135(2) | 2.78 | 3.32 |
| | 0.20 as-synthesized | 3.8797(1) | – | – | 58.398(3) | 2.82 | 3.30 |
| | 0.10 O$_2$-annealed | 3.8648(1) | – | – | 57.727(2) | 2.76 | 3.35 |
| | 0.15 O$_2$-annealed | 3.8714(1) | – | – | 58.025(3) | 2.78 | 3.32 |
| | 0.20 O$_2$-annealed | 3.8776(1) | – | – | 58.303(2) | 2.83 | 3.33 |
| Ru-substituted ($x$) | 0.05 as-synthesized | – | 3.8471(1) | 7.7232(2) | 57.153(5) | 2.74 | 3.40/3.45* |
| | 0.10 as-synthesized | – | 3.8544(1) | 7.7426(1) | 57.513(3) | 2.76 | 3.36/3.47* |
| | 0.15 as-synthesized | 3.8708(2) | – | – | 57.995(4) | 2.78 | 3.31/3.48* |
| | 0.20 as-synthesized | 3.8753(1) | – | – | 58.199(2) | 2.80 | 3.25/3.50* |
| | 0.10 O$_2$-annealed | – | 3.8506(1) | 7.7170(2) | 57.210(2) | 2.87 | 3.59/3.70* |
| | 0.15 O$_2$-annealed | 3.8641(1) | – | – | 57.696(2) | 2.87 | 3.52/3.69* |
| | 0.20 O$_2$-annealed | 3.8738(1) | – | – | 58.131(2) | 2.87 | 3.42/3.67* |

* The $V_{Co}$ values assuming that Ru is tetravalent (right) and pentavalent (left) in the structure.

Table II:
Transition temperature ($T_m$) and the magnetization ($M$) value in 5 T at 5 K of the SrCo$_{1-x}$Nb$_x$O$_{3-\delta}$ and SrCo$_{1-x}$Ru$_x$O$_{3-\delta}$ samples.

| | Sample | $T_m$ (K) | $M$ ($\mu_B$ / f.u.) |
|---|---|---|---|
| Nb-substituted ($x$) | 0.05 as-synthesized | 145 | 0.73 |
| | 0.10 as-synthesized | 130 | 0.30 |
| | 0.15 as-synthesized | 130 | 0.15 |
| | 0.20 as-synthesized | 130 | 0.12 |
| | 0.10 O$_2$-annealed | 150 | 0.53 |
| | 0.15 O$_2$-annealed | 150 | 0.28 |
| | 0.20 O$_2$-annealed | 150 | 0.18 |
| Ru-substituted ($x$) | 0.05 as-synthesized | 145 | 0.70 |
| | 0.10 as-synthesized | 130 | 0.37 |
| | 0.15 as-synthesized | 145 | 0.17 |
| | 0.20 as-synthesized | 165 | 0.17 |
| | 0.10 O$_2$-annealed | 165 | 1.18 |
| | 0.15 O$_2$-annealed | 180 | 0.54 |
| | 0.20 O$_2$-annealed | 175 | 0.23 |

Figure captions

Fig. 1:

X-ray diffraction pattern of the as-synthesized $SrCo_{0.90}Ru_{0.10}O_{3-\delta}$ sample. The pattern is indexed with a tetragonal "$a_p \times a_p \times 2\ a_p$" subcell. Extra reflections due to a tetragonal distortion are marked with a superscript "s". The inset shows [010] ED pattern of the same sample. The white arrows are associated with the presence of 90°-oriented twinned domains.

Fig. 2:

Temperature dependence of magnetization ($M$) for the $SrCo_{1-x}M_xO_{3-\delta}$ samples: (a) as-synthesized $M$ = Nb, (b) $O_2$-annealed $M$ = Nb, (c) as-synthesized $M$ = Ru, and (d) $O_2$-annealed $M$ = Ru. The data were recorded in an applying field of 0.3 T in the zero-field-cooling (solid symbols) and field-cooling (open symbols) modes using a SQUID magnetometer.

Fig. 3:

The $M$ vs $H$ loops at 5 K for the as-synthesized and $O_2$-annealed $SrCo_{1-x}Nb_xO_{3-\delta}$ samples with various Nb contents ($x$).

Fig. 4:

Temperature dependence of ac-susceptibility, (a) $\chi'$ and (b) $\chi''$, for the $SrCo_{1-x}Nb_xO_{3-\delta}$ samples. The data were recorded in an ac-field of 3 Oe with a frequency of 1 Hz. Open and closed symbols denote the data plots for the as-synthesized and $O_2$-annealed samples, respectively.

Fig. 5:

The ac-susceptibility, $\chi'$ and $\chi''$, of the $O_2$-annealed $SrCo_{0.90}Nb_{0.10}O_{3-\delta}$ (left) and $SrCo_{0.90}Ru_{0.10}O_{3-\delta}$ (right) samples. The data were recorded in an ac-field of 3 Oe with frequencies of 1, 10, and 100 Hz.

Fig. 6:

Temperature dependence of resistivity ($\rho$) for the SrCo$_{1-x}$M$_x$O$_{3-\delta}$ samples: (a) as-synthesized $M$ = Nb, (b) O$_2$-annealed $M$ = Nb, (c) as-synthesized $M$ = Ru, and (d) O$_2$-annealed $M$ = Ru. Solid and broken curves show $\rho$ data recorded in 0 T and 7 T, respectively. The transition temperature $T_m$ of each sample is marked with an arrow.

Fig. 7:

The comparison of MR vs $H$ curves (at 5 K) for the Nb- and Ru-substituted samples possessing the same substitution level ($x$, $y$ = 0.15, O$_2$-annealed). Solid and broken curves show MR data of the Nb- and Ru-substituted samples, respectively. In each MR-$H$ curves, the samples were zero-field cooled, and then the applied field was increased from 0 to 7 T and decreased back to 0 T, and further to -7 T, as guided with arrows in the figure.

Fig. 8:

The MR values (at 5 K, 7 T) against the Nb ($x$) and Ru ($y$) contents. The MR values were obtained from the MR vs $H$ measurements. Open and closed symbols denote the as-synthesized and O$_2$-annealed samples, respectively.

Fig. 9:

(a) The Curie temperature $T_C$ or magnetic transition temperature $T_m$, and (b) magnetization ($M$) value at 5 K in 5 T, for the SrCo$_{1-x}$Nb$_x$O$_{3-\delta}$ and SrCo$_{1-y}$Ru$_y$O$_{3-\delta}$ samples. These values are plotted against the cobalt valency, $V_{Co}$. For the SrCo$_{1-y}$Ru$_y$O$_{3-\delta}$ sample series, the $V_{Co}$ values were obtained based on the chemical formula assuming that all the substituted Ru ions are pentavalent in the structure. The data of the non-substituted SrCoO$_{3-\delta}$ phase reported by Taguchi *et al.* [2], Beddicka *et al.* [6], and Kawasaki *et al.* [7] are also plotted in the figures. Broken lines are guides to eye.

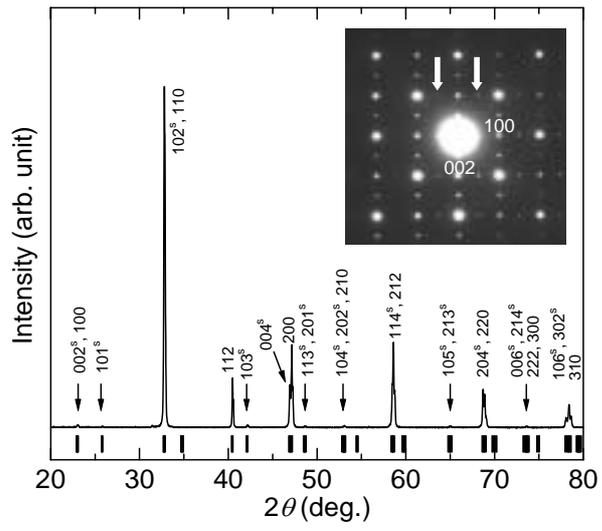

Fig. 1: Motohashi *et al.*

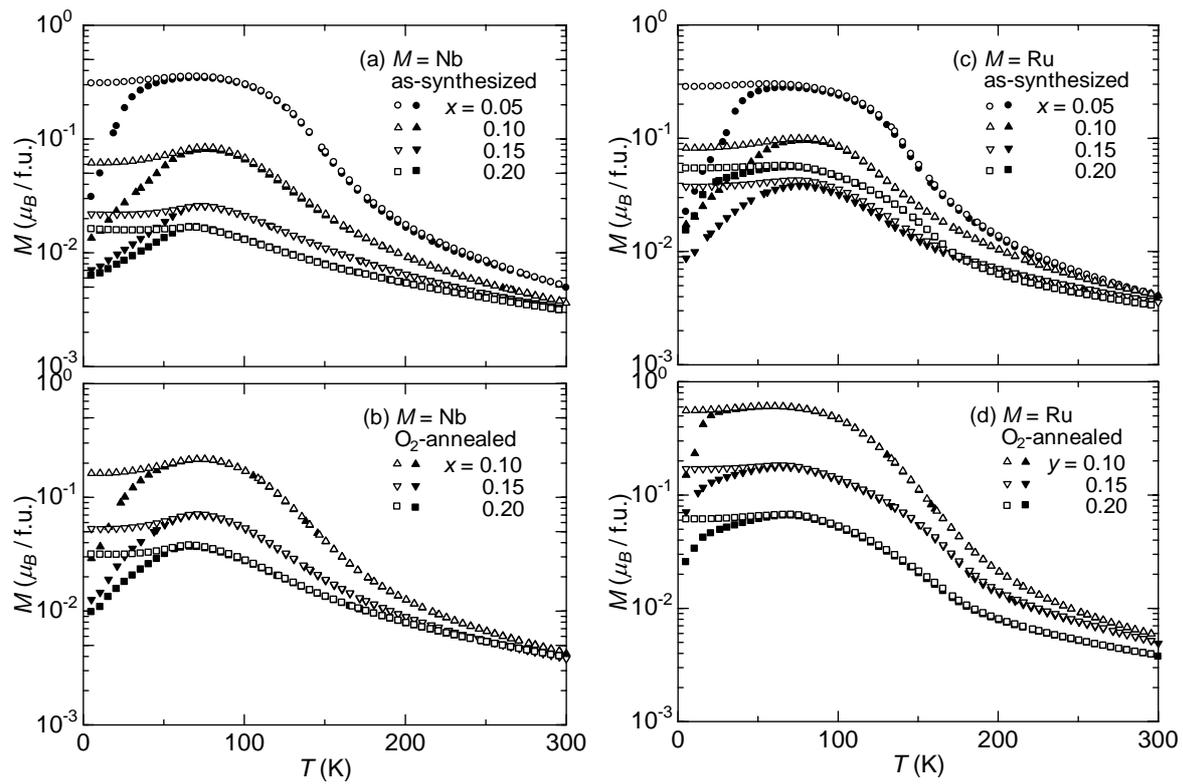

Fig. 2: Motohashi *et al.*

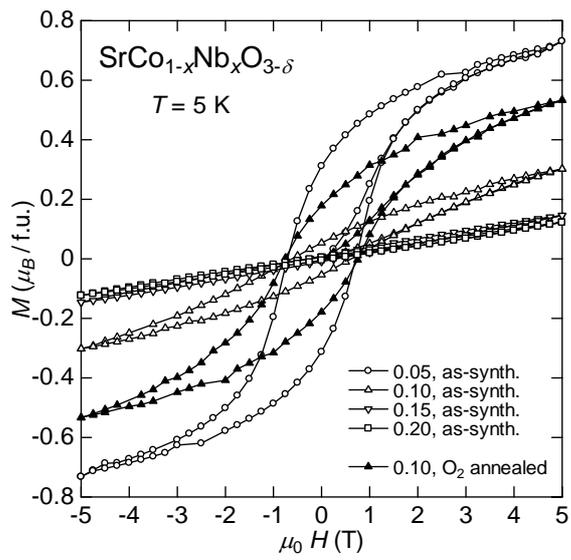

Fig. 3: Motohashi *et al.*

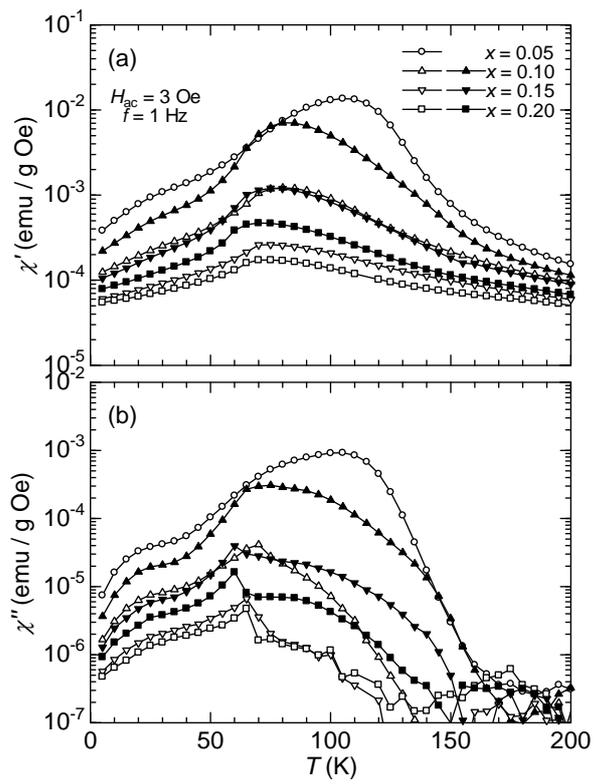

Fig. 4: Motohashi *et al.*

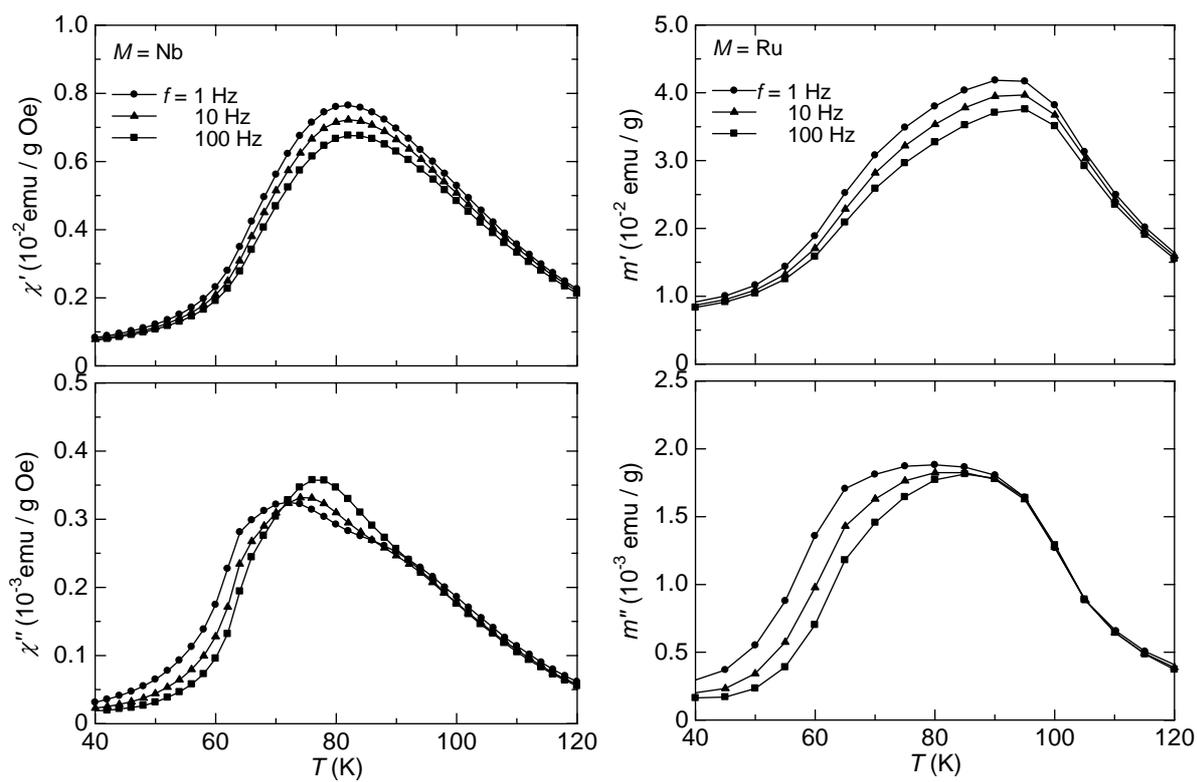

Fig. 5: Motohashi *et al.*

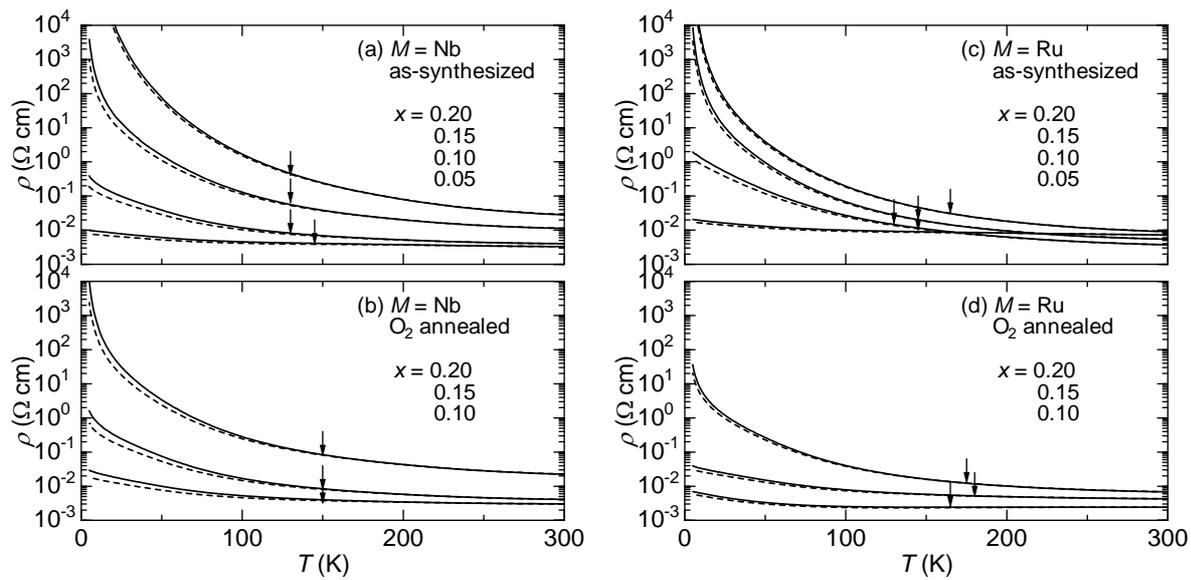

Fig. 6: Motohashi *et al.*

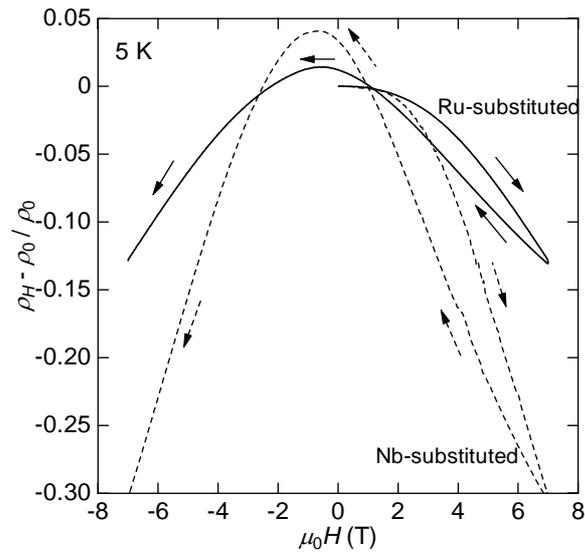

Fig. 7: Motohashi *et al.*

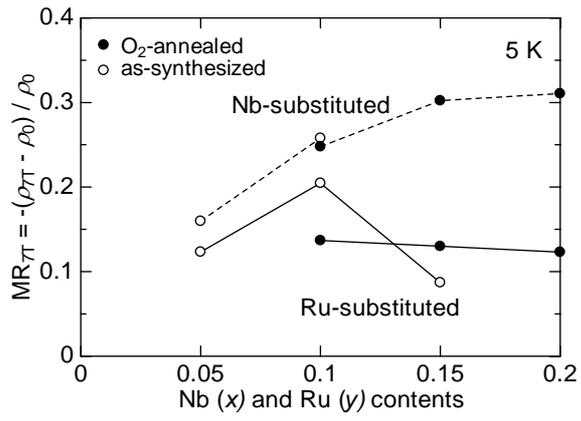

Fig. 8: Motohashi *et al.*

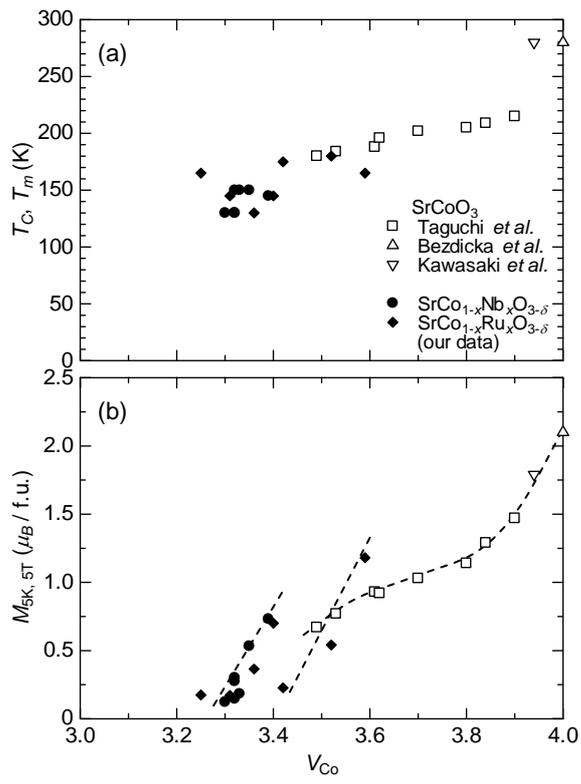

Fig. 9: Motohashi *et al.*